\documentclass[journal=ancac3,manuscript=article]{achemso}

\usepackage{chemformula} 
\usepackage{siunitx}
\usepackage{gensymb} 
\usepackage[version=4]{mhchem}
\usepackage[T1]{fontenc} 
\usepackage{float}
\usepackage{amssymb}
\usepackage{pdfpages}

\usepackage{tabularx}
\usepackage{enumitem}
\usepackage{soul}

\author{Pavel V. Kolesnichenko}
\affiliation[Heidelberg University]{Institute of Physical Chemistry, Heidelberg University, 69120, Heidelberg, Germany}
\email{pavel.kolesnichenko@alumni.uni-heidelberg.de}
\alsoaffiliation[CAM]{Institute for Molecular Systems, Engineering and Advanced Materials, Heidelberg University, 69120, Heidelberg, Germany}

\author{Lukas Wittenbecher}
\affiliation[Lund uni]{Department of Physics, Lund University, Box 118, 221 00 Lund, Sweden}
\affiliation[Lund University]{Division of Chemical Physics, Lund University, P.O. Box 124, 221 00 Lund, Sweden}
\affiliation[NanoLund]{NanoLund, P.O. Box 124, 221 00 Lund, Sweden}

\author{Qianhui Zhang}
\affiliation[Monash University]{Department of Civil Engineering, Monash University, Melbourne, Victoria 3800, Australia}

\author{Run Yan Teh}
\affiliation[Swinburne University]{Centre for Quantum Science and Technology Theory, Swinburne University of Technology, Melbourne, Victoria 3122, Australia}

\author{Chandni Babu}
\affiliation[Lund University]{Division of Chemical Physics, Lund University, P.O. Box 124, 221 00 Lund, Sweden}
\affiliation[NanoLund]{NanoLund, P.O. Box 124, 221 00 Lund, Sweden}

\author{Michael S. Fuhrer}
\affiliation[Monash University]{School of Physics and Astronomy, Monash University, Melbourne, Victoria 3800, Australia}
\alsoaffiliation{ARC Centre of Excellence in Future Low-Energy Electronics Technologies, Monash University, Melbourne, Victoria 3800 Australia}

\author{Anders Mikkelsen}
\affiliation[Lund uni]{Department of Physics, Lund University, Box 118, 221 00 Lund, Sweden}
\affiliation[NanoLund]{NanoLund, P.O. Box 124, 221 00 Lund, Sweden}

\author{Donatas Zigmantas}
\affiliation[Lund University]{Division of Chemical Physics, Lund University, P.O. Box 124, 221 00 Lund, Sweden}
\affiliation[NanoLund]{NanoLund, P.O. Box 124, 221 00 Lund, Sweden}
\email{donatas.zigmantas@chemphys.lu.se}

\title{Sub-100-fs formation of dark excitons in monolayer WS$_2$}

\keywords{WS$_2$, monolayer, excitons, 2D materials, transition metal dichalcogenide, intervalley scattering, femtosecond, microscopy, PEEM, pump-probe, semiconductor}

\begin{document}
\clearpage


\begin{abstract}


Two-dimensional semiconducting transition metal dichalcogenides (TMDs) are promising for optoelectronic applications due to their strongly bound excitons. While bright excitons have been thoroughly scrutinized, dark excitons are much less investigated as they are not observable with far-field spectroscopy. However, with their non-zero momenta, dark excitons are significant for applications requiring long-range transport or coupling to external fields. We access such dark excitons in WS$_2$ monolayer using transient photoemission electron microscopy with sub-diffraction limited spatial resolution (75 nm) and exceptionally high temporal resolution (13 fs). Image time series of TMD flakes are recorded at several different fluences. We directly observe the ultrafast formation of dark K-$\Lambda$ excitons in monolayer WS$_2$ occurring within 14–50 fs and follow their subsequent picosecond decay. We distinguish exciton dynamics between the interior and edges of the monolayer TMD and conclude that the long-term evolution of dark excitations is defect-mediated while intervalley scattering is not affected.

\end{abstract}

\section{Introduction}

Two-dimensional (2D) semiconductors such as monolayers of transition metal dichalcogenide (TMdC) are promising for applications in optics and optoelectronics because they exhibit rich exciton physics\cite{Mueller2018}. The large exciton binding energies of $\sim$0.5~eV in these materials make 2D excitons stable at room temperature, allowing for high exciton densities\cite{Chernikov2015}. This has enabled observations of various exciton formations including neutral\cite{Splendiani2010} and charged\cite{Mak2012} excitons, bi-excitons\cite{You2015}, and higher-order exciton complexes\cite{Hao2017,Zinkiewicz2021}. This, in turn, renders TMdC monolayers an ideal platform for investigating various many-body interactions and related emerging phenomena\cite{Sie2015,Kogar2017,Sun2021,Li2021,Xiao2021}. 

Excitons with trivial (zero) momentum form within TMdC monolayers in K valleys of their Brillouin zone, facilitated by enhanced electron-hole Coulomb interactions\cite{Chernikov2014}. These excitons have been thoroughly investigated in the literature\cite{Wang2018a} because they form within the light cone in energy-momentum space enabling their straightforward spectroscopic interrogation. Subsequent scattering of carriers to adjacent valleys leads to the formation of indirect excitons with nontrivial momenta falling outside the light cone\cite{Wallauer2021}, rendering them optically dark. Due to their net non-zero momentum, studying dark excitons is of importance because they can be better alternatives to bright excitons for long-range transport\cite{Sun2014,Rosati2021,Su2022,Lee2022,Lin2023,Chand2023,Katzer2023} or coupling to external fields such as those induced by plasmons\cite{Nerl2017,Lo2022}. Accessing nontrivial-momentum dark excitations, however, is an experimental challenge: optical spectroscopy cannot directly probe them, and momentum-resolved photoemission spectroscopy, although capable of directly detecting intervalley carriers, is experimentally highly challenging\cite{Zhang2022a}. Nevertheless, photoemission-based spectroscopies, remain essential for directly accessing momentum-forbidden dark excitons. 

Many studies have reported on the life-cycle of 2D excitons in various experimental settings covering their formation\cite{Ceballos2016,Steinleitner2017,Trovatello2020}, ultrafast cooling\cite{Li2020}, intervalley scattering\cite{Hein2016,Wang2018,Bao2020,Wallauer2021}, and ultimate fate (\textit{e.g.}, exciton-exciton annihilation\cite{Sun2014,Erkensten2021,Lee2022}, electron-hole recombination\cite{Lee2022}, and exciton dissociation\cite{Herman2022}). Exciton cooling, formation, and intervalley scattering are among the initial processes occurring on femtosecond timescales\cite{Reding2021} with the fastest reported being on a sub-100~\textit{fs} time scale\cite{Hein2016,Wang2018,Trovatello2020}. 
Therefore, in studying dark excitons, an additional challenge of resolving early-stage dynamics must be overcome -- advanced ultrafast spectroscopy techniques with very high temporal resolution are required. Optimizing the temporal resolution becomes even more crucial in cases when exciton dynamics occur on a sub-100-\textit{fs} time scale. However, pushing the resolution of femtosecond photoemission-based apparati to the level needed for confident identification of sub-100-\textit{fs} ultrafast processes poses a formidable challenge. 

Previously, femtosecond carrier kinetics in TMdC materials have been investigated using optical and photoemission-based spectroscopies with time resolutions in the range of 20--200~\textit{fs} with the most common value being a few tens of femtoseconds\cite{Ceballos2016,Steinleitner2017,Trovatello2020,Wang2018,Huber2019,Li2020,Xu2020,Sass2021,Liang2021,Hein2016,Madeo2020,Wallauer2021,Helmrich2021,Bertoni2016,Dong2021}. In particular, sub-100-\textit{fs} carrier dynamics have previously been identified in MoS$_2$\cite{Trovatello2020}, WSe$_2$\cite{Steinleitner2017,Wang2018}, and WS$_2$\cite{Wallauer2021} monolayers. In the latter case, an intervalley transfer time of $16\pm5$~\textit{fs} was observed from temporal offsets of momentum-resolved signals via two-photon photoemission with $>50$~\textit{fs} temporal resolution. In this study, we pushed the resolution below 20~\textit{fs} enabling straightforward observation in the temporal domain of sub-100~\textit{fs} formation of momentum-forbidden dark excitons in monolayer WS$_2$ measured in a simpler setting via one-photon photoemission.


 More specifically, we combine photoemission electron microscopy (PEEM) with high spatial resolution\cite{Maarsell2015,Wittenbecher2021,Vogelsang2021} (75~\textit{nm}) and femtosecond pump-probe spectroscopy with exceptionally high temporal resolution (13~\textit{fs}). By pumping at the main exciton resonance (2~\textit{eV}), we study subsequent carrier dynamics in WS$_2$ monolayer using the developed transient PEEM (TR-PEEM) apparatus. By design of the experiment, we detect ultrafast intervalley scattering from K valleys (\textit{i.e.}, the formation of dark excitons) occurring with a time constant in the range of $14-50$~\textit{fs}, detected via photoemission stimulated by broadband probe pulses in the deep-ultraviolet (DUV, 4.7~\textit{eV}). High temporal resolution was ultimately achieved by generating very short DUV pulses via achromatic phase matching in a nonlinear crystal, a method described previously\cite{Baum2004,Bruder2021} but never used before as part of transient photoemission microscopes. We finally take advantage of the imaging capabilities of the setup to additionally distinguish spatially-heterogeneous signals from the monolayer interior and edges, pointing at long-term defect-mediated processes.

\section{Results and discussion}

\begin{figure}[h!]
\centering
\includegraphics[width=0.9\linewidth,height=\textheight,keepaspectratio]{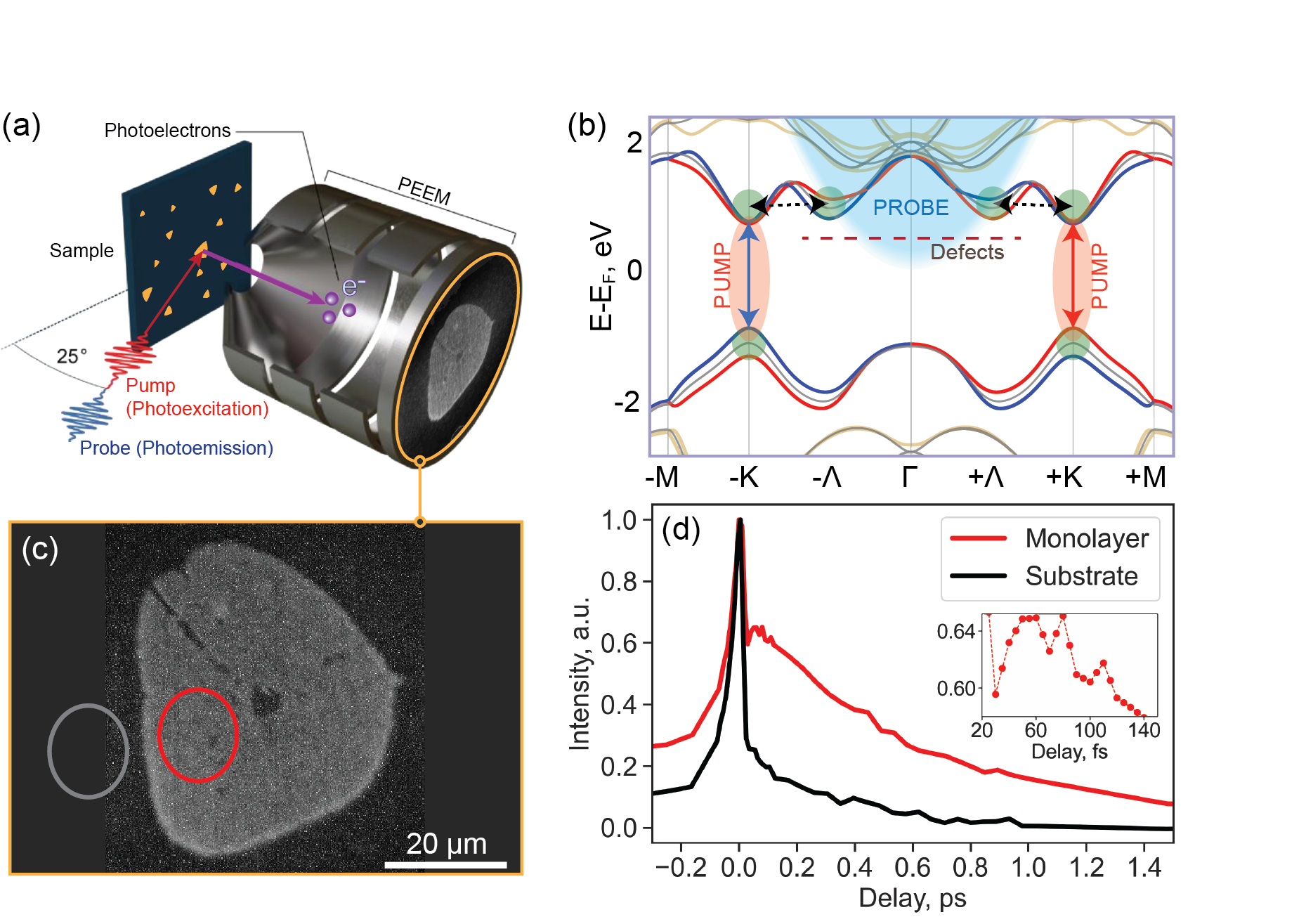}
\caption{(a) A TR-PEEM experiment. Pump pulse (red) excites carriers in monolayer WS$_2$ (orange); probe pulse (blue) photoemits electrons from the monolayer; these electrons form an image (encircled with orange line) for each pump-probe delay.  (b) Band structure of WS$_2$ monolayer calculated using DFT (Quantum Espresso\cite{Giannozzi2009}). Exciton transitions are indicated by blue and red arrows. Intervalley scattering is represented by dashed arrows. Area above photoemission horizon is schematically shown shaded in blue with a Gaussian edge reflecting final duration of DUV pulses. (c) An image of a WS$_2$ monolayer flake obtained using a Hg (mercury) discharge lamp. The length of the scale-bar is 20~$\mu{m}$. Grey and red circles indicate areas, across which signals in (d) were integrated for each pump-probe delay. (d) Normalized pump-probe traces spatially-integrated across the monolayer-interior (red) and substrate (black) regions shown in (c). The inset shows sub-100-fs rise and decay of photoemission signal. Pump and probe power densities were 136~$n{J}/cm^2$ and 70~$nJ/cm^2$ per pulse, respectively, and integration time was 10 sec.
}
\label{fig:fig1}
\end{figure}

The experiment is shown in Figure~\ref{fig:fig1}a (see also Methods, for more details). Prior to measuring carrier dynamics in the WS$_2$ monolayer, we first verified that the intensity of the pump beam was low enough to avoid high-order pump-induced photoemission from the flake (see Supplementary material, Section~S3). For the pump fluences used in our experiments, excitation density was estimated to be of the order of $10^{11}$~$cm^{-2}$ which is three orders of magnitude lower than the Mott density\cite{Chernikov2015}. This additionally ensures that the dominant pump-generated carriers are indeed bound electron-hole pairs (excitons)\cite{Steinhoff2017}. The energy of the probe pulses ultimately determined the photoemission horizon\cite{Li2020} in the energy-momentum space (Figure~\ref{fig:fig1}b),
\begin{equation}
\begin{aligned}
E_{probe}-\chi-E_b = \frac{\hbar{k}^2}{2m_e},
\end{aligned}
\label{eqn:PE_horizon}
\end{equation}
beyond which no photoemission is possible. In Eq.~(\ref{eqn:PE_horizon}), $E_{probe}$ is the probe pulse energy, $\chi$ is the effective electron affinity of monolayer WS$_2$, $E_b$ is the exciton binding energy, $m_e$ is the mass of electron, and $k$ is the in-plane momentum. As seen in Figure~\ref{fig:fig1}b, the detectable photoemission signal can only originate from the region (shaded blue) energetically higher than the photoemission horizon covering momenta in the vicinity of the $\Gamma$ point of the Brillouin zone, so that the valleys at the $\pm\Lambda$ points, which are energetically close to the conduction band minima, can contribute photoemitted electrons via K-to-$\Lambda$ intervalley scattering. Such scattering is expected to be energetically favourable in WS$_2$ monolayers\cite{Christiansen2017,Malic2018,Bao2020,Wallauer2021} and enhanced in $n$-doped monolayers\cite{Pei2023} leading to efficient generation of dark excitons. Additionally, the imminent presence of S-vacancies on the flake results in mid-gap states 0.3--0.5~\textit{eV} below the conduction band\cite{Carozo2017,Gao2021a} that can also fall into the probe region.

A photoemission image of a typical WS$_2$ monolayer flake is shown in Figure~\ref{fig:fig1}c. Similar flakes have been characterized previously\cite{Kolesnichenko2020,Kolesnichenko2021,McCreary2016} revealing the energy of main exciton resonance to be $\sim$2~\textit{eV} and that of charged excitons (trions) to be a few tens of meV (a charging energy) below. In TR-PEEM experiments reported here, the energy of the pump pulses was tuned to overlap the main exciton resonance thus promoting adiabatic exciton formation\cite{Trovatello2020,Wallauer2021}. The broad spectrum of the pump pulses also overlapped with the energy of trions. The resulting coherent exciton polarization is expected to loose its coherence within 100~\textit{fs}\cite{Trovatello2020,Wallauer2021} giving rise to incoherent excitons and trions. Pump-induced dynamics were then monitored via probe-induced photoemission. By the construction of our experiment, TR-PEEM is sensitive to dark intervalley carriers inaccessible by conventional optical spectroscopies. Measured typical pump-probe traces from the substrate and the interior of the monolayer WS$_2$ are shown in Figure~\ref{fig:fig1}d. Similar traces were also observed from other flakes on the substrate. In contrast to the substrate, the monolayer WS$_2$ features a prominent well-resolved rise of the photoemission signal on a sub-100-\textit{fs} time scale. 

\begin{figure}[h!]
\centering
\includegraphics[width=0.8\linewidth,height=\textheight,keepaspectratio]{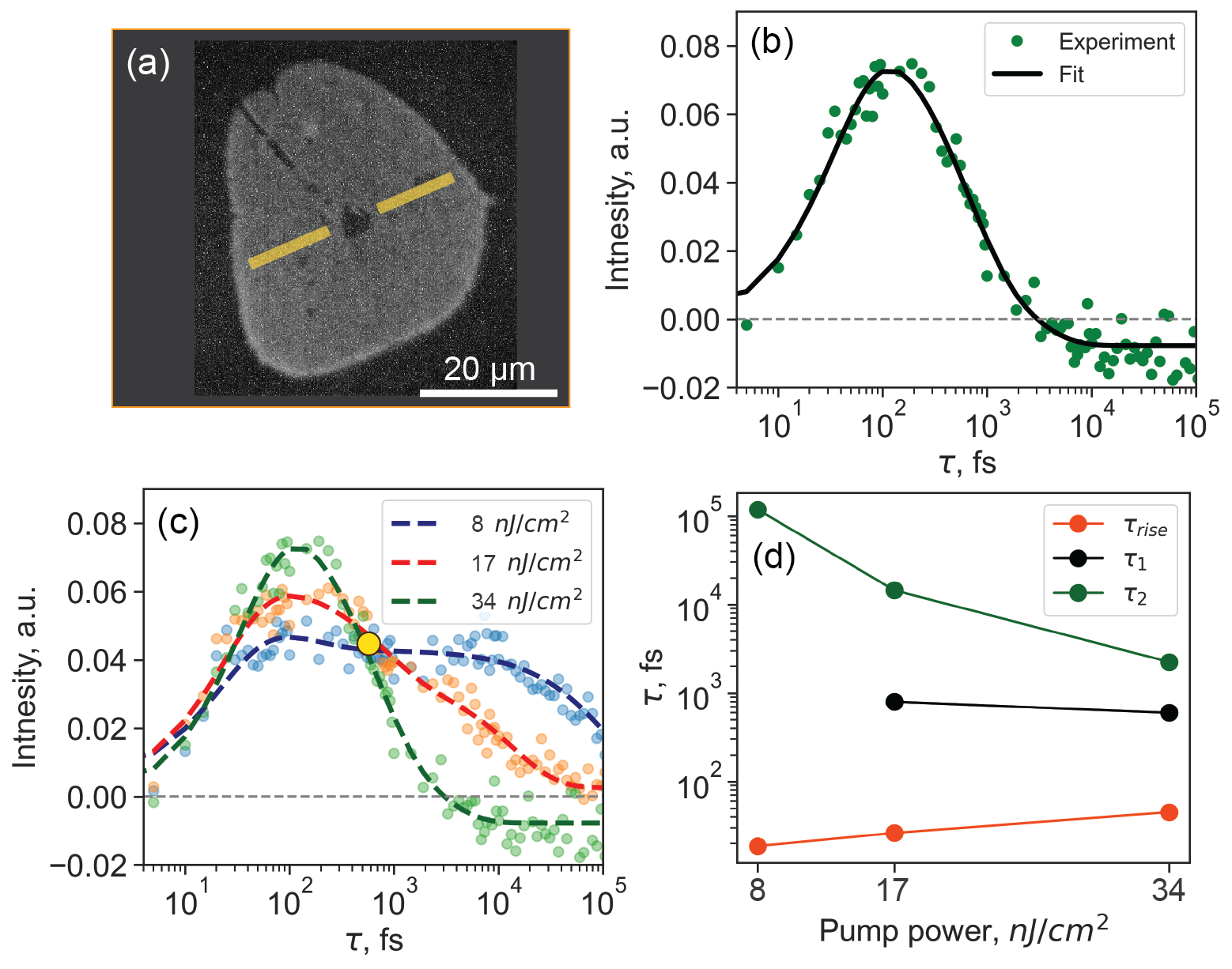}
\caption{(a) Area (yellow), across which signal was integrated for analysis. (b) Fitting of the pump-probe trace obtained by integrating the pump-probe signal across the area (indicated in (a)) on WS$_2$ flake. Pump fluence was 34~$n{J}/cm^2$. (c) Pump-probe traces (dots) and corresponding fits (dashes) from the same area as in (a) acquired for pump fluences of 8 (blue), 17 (red), and 34 (green) $nJ/cm^2$. Isometric point is indicated by yellow circle. (d) Power-dependence of extracted time-constants $\tau_{rise}$ (red), $\tau_1$ (black), and $\tau_2$ (green). The first $\tau_2$ point is not shown since it is not representative of the second rising signal.}
\label{fig:fig2}
\end{figure}

Before conducting a more detailed analysis of the observed carrier dynamics, it is necessary to assess whether the substrate contributed to the photoemission background in the measured pump-probe signals across the monolayer flake. The work function of Si and SiO$_2$ have been previously reported to be 4.8~\textit{eV} and 4.4~\textit{eV}, respectively\cite{Lindmayer1966}. Given the pump (2.0~\textit{eV}) and probe (4.7~\textit{eV}) energies used in this work, it is likely for substrate-electrons to be photoemitted from at least the SiO$_2$ layer with excess energy $E_e$ (electron kinetic energy) in the range of $\sim$0.3--2.3~eV. The de Broglie wavelength $\lambda_e$ of such electrons is estimated as $\lambda_e=h/\sqrt{2m_eE_e}$ ($m_e$ is electron's mass, and $h$ is Planck's constant) yielding the values in the range of 0.81--1.22~nm, which are larger than the thickness $\sim$0.6~nm of WS$_2$ monolayers\cite{Okada2019}. The inelastic mean free path of such electrons within the monolayer is expected to be greater than 80~nm in accordance with the universal curve for inorganic compounds\cite{Seah1979}, exceeding the monolayer's thickness by two orders of magnitude. Therefore, a substantial photoemission background from the substrate is indeed expected to contribute to the detected transient photoemission signals obtained from the monolayer region. We thus use photoemission from the substrate as an estimate for such background and subtract it from the overall photoemission signal in a manner that naturally suppresses coherent contributions in pump-probe traces (see Supplementary material, Section S4, for more details). This procedure also ensures that substrate effects such as surface space charge region, surface dipoles, surface carrier recombination, and surface state distribution\cite{Kronik1999} are also taken into account in the further analysis. The retrieved transient WS$_2$-specific photoemission contrast is shown in Figure~\ref{fig:fig2}. As expected, these differential dynamics feature a delayed rise of the photoemission signal followed by its subsequent decay. To gain more insights into the underlying carrier dynamics, we applied a simple fitting model to the resultant traces (see Supplementary material, Section S5), which takes into account the finite rise of the detected signal\cite{Singh2016,Trovatello2020},
\begin{equation}
\begin{aligned}
I = (1-e^{-\frac{\tau-\tau_0}{\tau_{rise}}})\cdot (a_0+a_1e^{-\frac{\tau-\tau_0}{\tau_{1}}}+a_2e^{-\frac{\tau-\tau_0}{\tau_{2}}}) \cdot{H}(\tau-\tau_0),
\end{aligned}
\label{eqn:fit_model}
\end{equation}
where $\tau_{rise}$ is the time-constant describing the sub-100-\textit{fs} photoemission build-up; $\tau_{1,2}$ describe subsequent dynamics beyond 100~\textit{fs}; $a_{1,2}$ are corresponding fitting amplitudes; $a_0$ and $\tau_0$ are constant photoemission offset and time-zero calibration, respectively.

Figure~\ref{fig:fig2}b shows the result of fitting applied to a pump-probe trace obtained with 34~$nJ/cm^2$ pump pulses, featuring a high fitting quality. We note a negative offset at longer delays indicating relative photoemission from the flake is less than that from the substrate, which we attribute to the effects of bandgap renormalization\cite{Wang2013,Pogna2016,Cunningham2017,Lin2019,Meckbach2020,Ataei2021}. Such renormalization could result in $\Lambda$-valleys lowering their energy below the photoemission horizon and/or less energetically-favourable scattering to the probe region (if K-valleys become lower in energy than $\Lambda$-valleys). Regardless of these two possible mechanisms, the net effect from both of them is a reduction in the photoemission signal for higher pump fluences. Therefore, to provide support for the hypothesis of the bandgap renormalization effects, we measured pump-induced photoemission for higher pump fluences resulting in an even weaker photoemission signal (see Supplementary material, Section S6). 

The retrieved time-constants are $\tau_{rise}\sim45$~fs, $\tau_1\sim0.6$~ps, and $\tau_2\sim2.2$~ps, which are similar to those of previously-reported processes in 2D TMdC-based semiconductors, namely, (\textit{rise}) exciton formation\cite{Wang2018,Trovatello2020} and K-to-$\Lambda$ intervalley scattering\cite{Wang2018,Wallauer2021}, and (\textit{1,2}) reversal $\Lambda$-to-K intervalley scattering\cite{Wang2018}, exciton decay\cite{Ceballos2016,Madeo2020}, exciton-exciton annihilation\cite{Wang2018}, and the formation of trions\cite{Singh2016}. 
Due to the very low excitation densities used in our experiments, we exclude exciton-exciton annihilation processes. Geminate excitons form outside the probe region in the energy-momentum space and therefore should not contribute to photoemission either. In addition, their formation should be independent of pump fluence, but scattering processes are, in contrast, pump-fluence dependent. In the latter case, we note that for such low excitation densities used in this work, exciton-exciton scattering is not a plausible intervalley transfer mechanism; rather, it is scattering mediated by phonons\cite{Christiansen2017,Raja2018} and/or plasmons\cite{VanTuan2017} (due to intrinsic \textit{n}-doping of CVD-grown monolayers\cite{Chae2017}) that are likely at play in this case. To confirm the scattering processes taking place and to gain further insight into the detected sub-100-\textit{fs} photoemission rise and subsequent dynamics, we acquired pump-fluence dependences of pump-probe signals (Figure~\ref{fig:fig2}c,d).

Figure~\ref{fig:fig2}c shows pump-probe dynamics measured for three pump fluences of 8~$nJ/cm^2$, 17~$nJ/cm^2$, and  34~$nJ/cm^2$. Notably, fluence-dependent changes in amplitudes occur in opposite directions for delays shorter and longer than $\sim$700~\textit{fs}. Specifically, for delays $<700$~\textit{fs} photoemission intensity overall increases with pump fluence, whereas for delays $>700$~\textit{fs} photoemission intensity decreases. This behaviour indicates a possible interplay of competing processes, which is also supported by the observation of a second rising signal (peaking at $\sim$7~ps) most prominent in the pump-probe dynamics obtained with the lowest pump fluence of 8~$nJ/cm^2$ (see also Supplementary material, Section~S6). This observation points to a possible sequential carrier-transfer phenomenon that becomes more dominant for low pump fluences. The delay of $\sim$700~\textit{fs}, in this case, can be regarded as an isometric point\cite{Kaspar2023} in the temporal domain, \textit{i.e.}, the point at which the signal is independent of pump fluence (which is in agreement with nearly constant $\tau_1$ for higher pump fluences). We note that the chosen model (Eq.~(\ref{eqn:fit_model})) fits well the two higher-fluence pump-probe traces, but does not explicitly take into account this secondary carrier transfer observed in the low-fluence case. Instead, by applying a fitting model that takes into account two rising and decaying signals (see Supplementary material, Section~S7), we can obtain a better fit to the pump-probe trace for the lowest excitation density. In this case, signal rise times of $\tau_{rise,1}\sim23.7$~fs and $\tau_{rise,2}\sim1.1$~ps are retrieved, which are in agreement with the reported values of intervalley exciton scattering\cite{Wang2018,Wallauer2021} and trion formation\cite{Singh2016}, respectively. In Figure~\ref{fig:fig2}c, nevertheless, the fluence-dependent trends are extracted using the model described by Eq.~(\ref{eqn:fit_model}) for consistency.

All three time constants, overall, are fluence-dependent (Figure~\ref{fig:fig2}d) indicating contributions from scattering-mediated processes, with $\tau_{rise}$ increasing, and $\tau_2$ decreasing with pump fluence. Initially, the time-constant $\tau_1$ increases and then mildly decreases for increasing pump fluence (see Supplementary material, Section S5), which also supports an interplay of competing processes mentioned above. A lower scattering rate, as indicated by larger $\tau_{rise}$, for larger pump fluences further fortifies the notion that there are additional effects of bandgap renormalization at play. 
The decreasing trend for $\tau_2$ also additionally supports the above-mentioned second signal-rise wave, which in this case likely reflects, indirectly, backscattering from $\Lambda$ to $K$ valleys or, directly, decay to longer-lived defect states (such as those introduced by S-vacancies) and/or generation of trions. Further studies are required to disentangle all these effects. Given the discussion above, the time-constant $\tau_1$ reflects an interplay between exciton decay to lower states (such as traps and trions) within the probe region, and $\Lambda$-to-K carrier scattering away from the probe region. Thus, for lower fluences, it is the exciton decay to longer-lived states that is prevalent resulting in lower rates and a prominent second rise of photoemission. For higher fluences, when bandgap renormalization takes place, energetically-lowered K-valleys could result in a more efficient backscattering. 

\begin{figure}[h!]
\centering
\includegraphics[width=1\linewidth,height=\textheight,keepaspectratio]{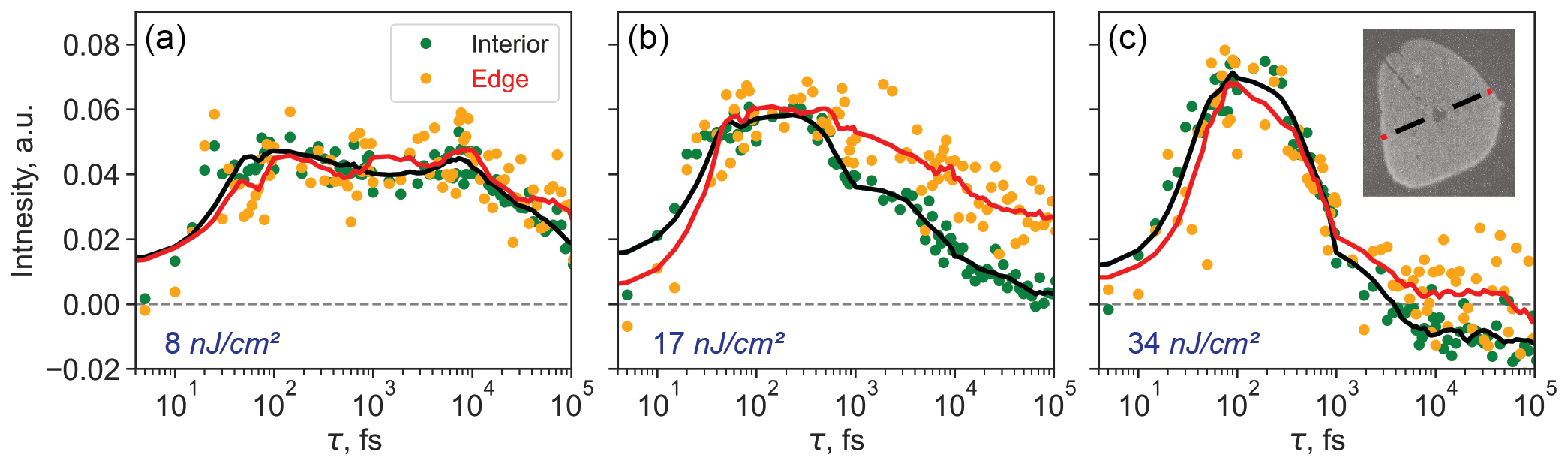}
\caption{Comparison of pump-probe signals acquired from interior (green dots and black lines) and edges (orange dots and red lines) of the WS$_2$ flake for three excitation powers of (a) 8, (b) 17. and (c) 34~$n{J}/cm^2$. Solid lines represent the result of smoothing to reveal fine details of the traces. Signal integration areas are indicated in inset in (c).}
\label{fig:fig3}
\end{figure}

We finally note that the measured carrier dynamics is most likely defect-mediated. It has been observed previously that edges in CVD-grown WS$_2$ monolayers contain larger amount of S-vacancies compared to the interior\cite{Bao2015,Carozo2017,Rosenberger2018}. Defect densities in the interior can be as large as $\sim$10$^{13}$~$cm^{-2}$, whereas those near the edges of the order of $\sim$10$^{14}$~$cm^{-2}$\cite{Carozo2017}, which are 2--3 orders of magnitude larger than the excitation densities in this work (see also Ref.\cite{Bao2015}). Therefore, we next take advantage of the high spatial resolution of the method, and directly compare pump-probe dynamics from the interior (with a lesser number of S-vacancies) and edges (with a larger density of S-vacancies) of the monolayer flake. Figure~\ref{fig:fig3} shows pump-probe dynamics from the interior and edges for the three investigated pump fluences. In all cases, a slower decay dynamics from the edges is evident confirming contributions from trap-mediated exciton dynamics across the flake. Notably, the intervalley scattering rates are not as different between the interior and edges of the monolayer (see Supplementary material, Section S5) indicating that defects likely do not act as efficient scattering centers in this case. Nevertheless, compared to the interior, at the edges there is a clear trend of overall lower sub-100-fs photoemission accompanied by larger photoemission on picosecond timescale (with photoemission intensity unchanged at sub-picosecond delays) similar to what was observed in fluence-dependent measurements. This further supports that the prevailing decay to longer-lived states for lower pump-fluences discussed above is indeed the decay to defect states. 

Therefore, given the overall discussion above, it can be concluded that, within the range of set experimental conditions, dark excitons in monolayer WS$_2$ form within $\sim$14--50 \textit{fs} followed by picosecond-scale dynamics mediated by defects. This ultrafast formation of dark excitons occurs as fast as the cooling of bright excitons\cite{Trovatello2020} suggesting that dark-exciton configuration is more energetically favourable in WS$_2$ monolayers.


\section{Conclusion}

In summary, we investigated intervalley carrier dynamics in monolayer WS$_2$ via PEEM coupled to femtosecond pump-probe spectroscopy with a very-high temporal resolution of 13~\textit{fs}. We identified initial K-to-$\Lambda$ intervalley scattering (formation of dark excitons) occurring on a time scale of 14--50~\textit{fs}, depending on the excitation and defect density. The intervalley scattering does not appear to differ significantly between the edges and the interior of WS$_2$ monolayer. Subsequent dynamics suggested a decay of dark excitons to longer-lived states. A longer defect-mediated dynamics at the monolayer edges in this case were unambiguously identified by taking advantage of the imaging capabilities of the apparatus with a sub-diffraction-limited spatial resolution of 75~\textit{nm}. The developed spectroscopy approach can be used for the direct identification of sub-100-\textit{fs} processes in other TMdC monolayers as well as graphene\cite{Trepalin2019} and other topological semimetals\cite{Biswal2022}. Furthermore, complementing the ultrafast TR-PEEM method with excitation frequency resolution\cite{Huber2019} and energy resolution of the photoemitted electrons will provide a more comprehensive picture of the ultrafast processes taking place in these materials.

\section{Methods}

\subsection{Sample preparation}

Monolayers of WS$_2$ were grown on sapphire (Al$_2$O$_3$) substrate via chemical vapor deposition (CVD) following a similar procedure described previously\cite{Zhang2018}. For TR-PEEM experiments, the monolayers were subsequently transferred onto an $n$-doped silicon wafer with a natural oxide layer (SiO$_2$/$n$-Si).

\subsection{Electronic-band-structure calculation}

Electronic band structure of WS$_2$ monolayer was calculated using density functional theory (DFT) as implemented in Quantum Espresso\cite{Giannozzi2009}.

\subsection{TR-PEEM experiment}

Broadband visible excitation (pump) pulses ($\sim$10~\textit{fs} duration, $\sim$2~\textit{eV} central energy, $\sim$320~\textit{meV} spectral bandwidth, 2.7--44~\textit{pJ} energy per pulse) were generated in a lab-built non-collinear optical parametric amplifier (NOPA). Broadband deep-UV ionization (probe) pulses ($\sim$10~\textit{fs} duration, $\sim$4.7~\textit{eV} central energy, $\sim$330~\textit{meV} spectral bandwidth, 7~\textit{pJ} energy per pulse) were generated as second harmonic of the output (with 2.34~\textit{eV} central energy) from a second NOPA via achromatic phase matching\cite{Baum2004,Bruder2021}. Pulse repetition rate was 100~kHz. Both the excitation and ionization beams were weakly focused onto the sample inside the PEEM vacuum chamber at an angle of 25$\degree$ with respect to the sample surface (Figure~\ref{fig:fig1}a). Both beams had ellipse-shaped spots on the sample surface with long and short axes being $\sim$130~\textit{µm} and $\sim$80~\textit{µm} (estimated fwhm) in the case of pump beam, and $\sim$200~\textit{µm} and $\sim$50~\textit{µm} in the case of probe beam, respectively. The pump and probe fluences at the sample were in the range of 8--136~$nJ/cm^2$ and 70~$nJ/cm^2$, respectively. The power in the probe beam was chosen low enough for samples not to degrade over the course of experiments\cite{Li2020} as well as for images to be acquired without space charge effects\cite{Buckanie2009}, but sufficiently high to be able to observe prominent pump-induced dynamics. Integration times during signal acquisition for each pump-probe delay were 40~\textit{sec} (for 8 and 17~$nJ/cm^2$ pump), 20~\textit{sec} (for 34~$nJ/cm^2$ pump), and 10~\textit{sec} (for 136~$nJ/cm^2$ pump). During TR-PEEM experiments, samples were contained in ultrahigh vacuum ($\sim$10$^{-10}$~$mbar$) inside a commercial PEEM apparatus (IS-PEEM, Focus GmbH) where electrons photoemitted from the sample ultimately formed an image on a charge-coupled device (CCD). The temporal and spatial resolution of the setup was estimated to be $\sim$75~\textit{nm} and $\sim$13~\textit{fs}, respectively  (see Supplementary material, Sections S1,S2).

\medskip
\textbf{Acknowledgements} \par 
The work was supported by Vetenskapsrådet, Crafoordska Stiftelsen, and NanoLund. M.S.F. acknowledges support from the Australian Research Council (DP200101345 and CE170100039).

\textbf{Authors disclosure statement} \par 
Authors declare no competing financial interests.

\begin{suppinfo}


The following file is available as Supplementary material:
\begin{itemize}
  \item supplementary.pdf
\end{itemize}

Pulse spectra and time-resolution of TR-PEEM; Spatial resolution of TR-PEEM; Pump-induced excitation and photoemission; Monolayer-specific dynamics; Fittings of pump-probe traces; Photoemission contrast drop for higher pump fluences; Fitting with two rise-and-decay signals (PDF)

\end{suppinfo}

\medskip
\bibliography{main.bib}




\section*{Supplementary material}

# Supplementary material to: "Sub-100-fs formation of dark excitons in monolayer WS$_2$"

Pavel V. Kolesnichenko,[*,†,‡] Lukas Wittenbecher,[¶] Qianhui Zhang,[⊥] Run Yan Teh,[#] Chandni Babu,[§] Michael S. Fuhrer,[@,△] Anders Mikkelsen,[¶] and Donatas Zigmantas[*,§]

[†]Institute of Physical Chemistry, Heidelberg University, 69120, Heidelberg, Germany

[‡]Institute for Molecular Systems, Engineering and Advanced Materials, Heidelberg University, 69120, Heidelberg, Germany

[¶]Department of Physics, Lund University, Box 118, 221 00 Lund, Sweden

[§]Division of Chemical Physics, Lund University, P.O. Box 124, 221 00 Lund, Sweden

[∥]NanoLund, P.O. Box 124, 221 00 Lund, Sweden

[⊥]Department of Civil Engineering, Monash University, Melbourne, Victoria 3800, Australia

[#]Centre for Quantum Science and Technology Theory, Swinburne University of Technology, Melbourne, Victoria 3122, Australia

[@]School of Physics and Astronomy, Monash University, Melbourne, Victoria 3800, Australia

[△]ARC Centre of Excellence in Future Low-Energy Electronics Technologies, Monash University, Melbourne, Victoria 3800 Australia

E-mail: pavel.kolesnichenko@alumni.uni-heidelberg.de; donatas.zigmantas@chemphys.lu.se

# S1. Pulse spectra and time-resolution of TR-PEEM

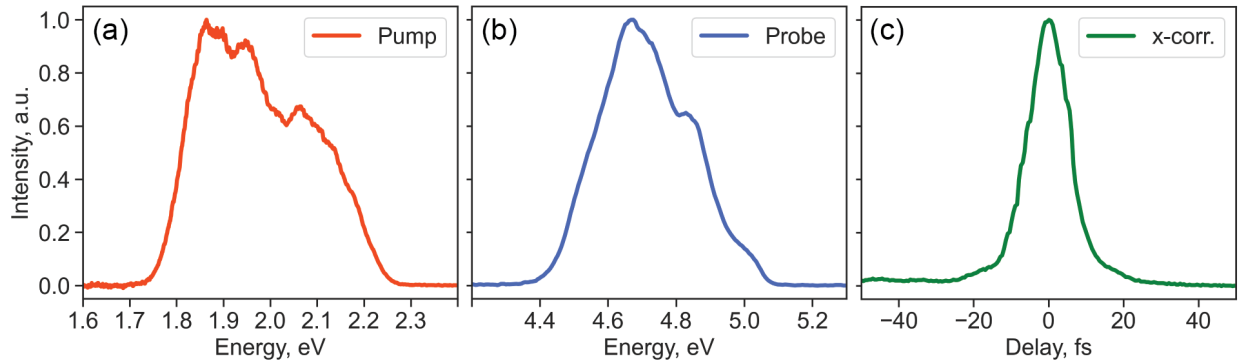

Figure S1: (a) Pump and (b) probe spectra, and (c) temporal cross-correlation (x-corr.) of pump and probe pulses measured via pump-induced transient-grating gate in a thin fused silica plate.[1] The x-corr. was measured in front of the PEEM's input window taking the window material into account. The fwhm of x-corr. profile is 13.4 $fs$ defining temporal resolution of the TR-PEEM apparatus.

# S2. Spatial resolution of TR-PEEM

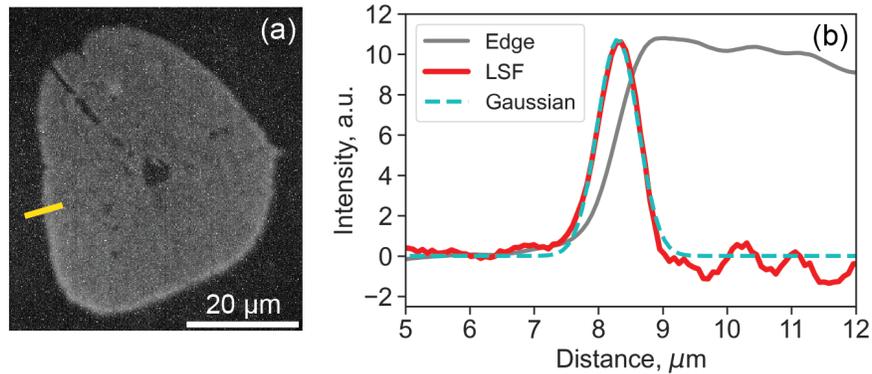

Figure S2: (a) Monolayer $WS_2$ flake imaged via a mercury lamp. (b) Edge profile (grey) taken along the yellow line indicated in (a), line spread function (LSF, red) of the edge profile, and a Gaussian fit to LSF with fwhm of 74.6 $nm$ defining spatial resolution of the TR-PEEM apparatus.

# S3. Pump-induced excitation and photoemission

Table S1: Pump fluences and corresponding carrier excitation densities.

| Pump fluence, $nJ/cm^2$ | 8 | 17 | 34 | 136 |
|---|---|---|---|---|
| Excitation density, $10^{11} \times cm^{-2}$ | 0.52 | 1.05 | 2.12 | 8.49 |

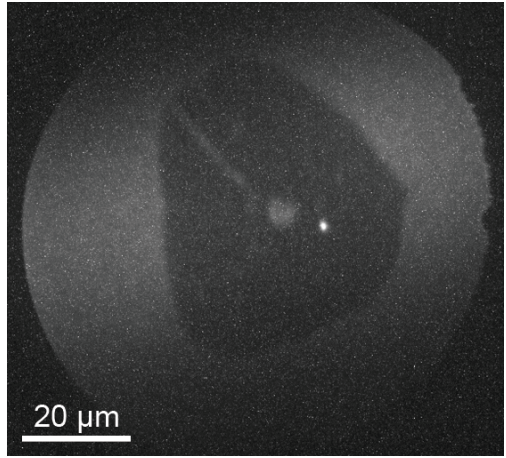

Figure S3: A photoemission image of the WS$_2$ monolayer flake obtained using 136 $nJ/cm^{-2}$ pump pulses (with probe pulses blocked).

# S4. Monolayer-specific dynamics

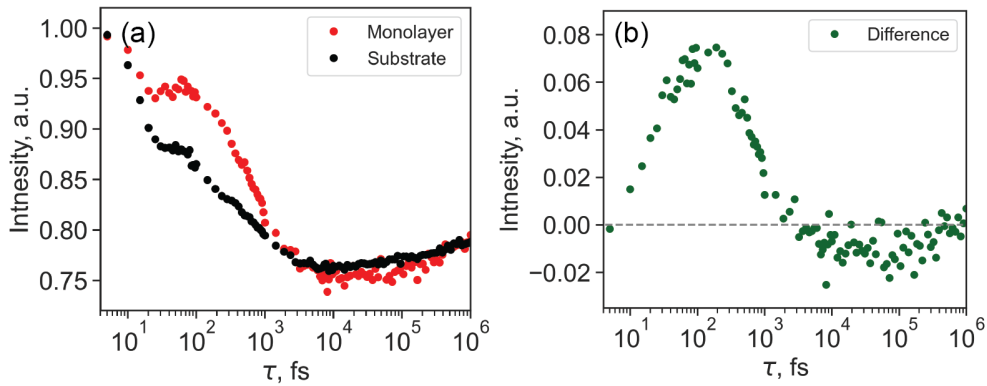

Figure S4: (a) Normalization of raw pump-probe traces obtained from WS$_2$ monolayer interior and substrate. (b) Difference (also Figure 2a of the main text) between the pump-probe traces in (a). Pump fluence was 34 $nJ/cm^2$.

The two spectra were pinned at 0 fs and 1 ns delays (Figure S4a) and subtracted (Figure S4b. This procedure naturally suppresses the effects from pulse overlap including coherent exciton population signal.

## S5. Fittings of pump-probe traces

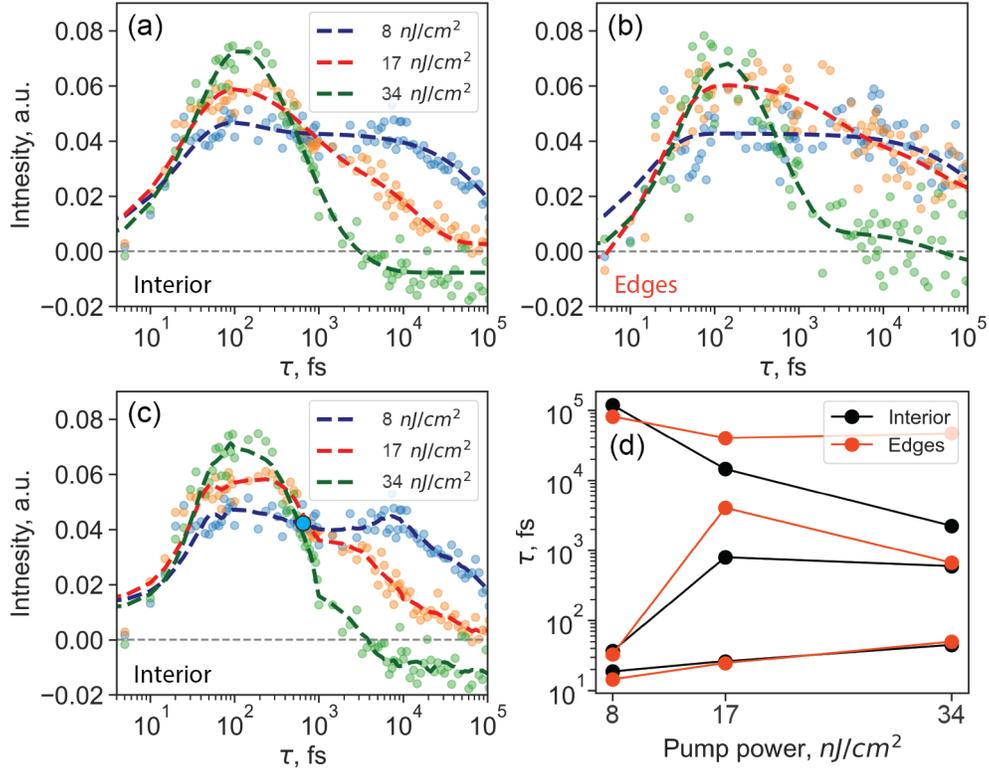

Figure S5: Fittings of the pump-probe traces acquired from the interior (a) and edges (b) of the monolayer flake for three pump fluences. (c) A result of smoothing of pump-probe traces revealing fine details such as a secondary rising signal. (d) Dependence of extracted time constants on pump fluence.

Table S2: Extracted time-constants for the three pump fluences (interior of the flake)

|  | $8\ nJ/cm^2$ | $17\ nJ/cm^2$ | $34\ nJ/cm^2$ |
|---|---|---|---|
| $\tau_{rise}$, $fs$ | 18.7 | 26.3 | 45.1 |
| $\tau_1$, $fs$ | 36.9 | 797 | 596 |
| $\tau_2$, $ps$ | 118.2 | 14.5 | 2.2 |

Table S3: Extracted time-constants for the three pump fluences (flake's edges)

|  | $8\ nJ/cm^2$ | $17\ nJ/cm^2$ | $34\ nJ/cm^2$ |
| --- | --- | --- | --- |
| $\tau_{rise},\ fs$ | 14.5 | 24.6 | 50.0 |
| $\tau_1,\ fs$ | 32.7 | 2747 | 624 |
| $\tau_2,\ ps$ | 81.6 | 40.2 | 46.6 |

# S6. Photoemission contrast drop for higher pump fluences

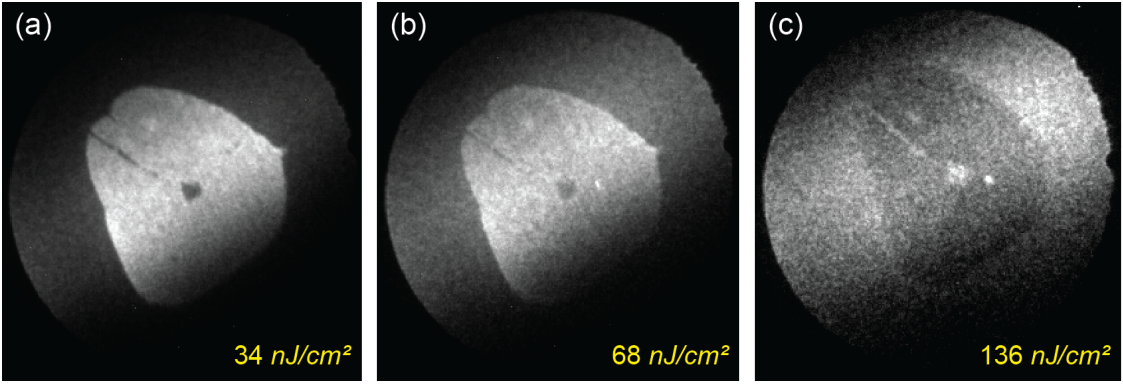

Figure S6: Photoemission at 7 ps pump-probe delay for three pump fluences of (a) $34\ nJ/cm^2$, (b) $68\ nJ/cm^2$, and (c) $136\ nJ/cm^2$. For higher fluences photoemission contrast from the flake becomes lower.

# S7. Fitting with two rise-and-decay signals

Extracted rise time-constants are $\tau_{rise,1} \sim 23.7$ fs and $\tau_{rise,2} \sim 1.1$ ps comparable to previously reported values of dark-exciton[2,3] and trion formation.[4]

$$I = \left( \left(1 - a_{rise,1} e^{-\frac{\tau - \tau_{0,1}}{\tau_{rise,1}}}\right) \cdot a_1 e^{-\frac{\tau - \tau_{0,1}}{\tau_1}} + \left(1 - a_{rise,2} e^{-\frac{\tau - \tau_{0,2}}{\tau_{rise,2}}}\right) \cdot a_2 e^{-\frac{\tau - \tau_{0,2}}{\tau_2}} + a_0 \right) \cdot H(\tau - \tau_0)$$
(S1)

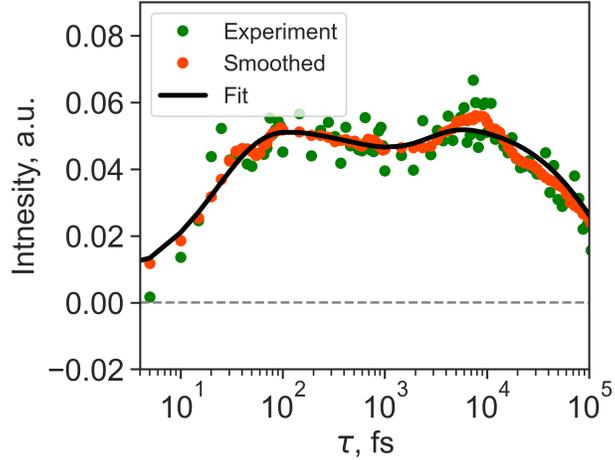

Figure S7: Fitting with a model (Eq. (S1)) accounting for two rise-and-decay signals. Pump fluence is 8 $nJ/cm^2$.



\end{document}